\documentclass[10pt,conference]{IEEEtran}
\IEEEoverridecommandlockouts

\usepackage{cite}
\usepackage{amsmath,amssymb,amsfonts}
\usepackage{algorithmic}
\usepackage{graphicx}
\usepackage{textcomp}
\usepackage{xcolor}
\usepackage{marvosym}
\usepackage{multirow}
\usepackage{makecell}
\usepackage{booktabs}

\def\BibTeX{{\rm B\kern-.05em{\sc i\kern-.025em b}\kern-.08em
    T\kern-.1667em\lower.7ex\hbox{E}\kern-.125emX}}
\begin{document}

\title{SGCR: A Specification-Grounded Framework for Trustworthy LLM Code Review
}


\author{
\IEEEauthorblockN{
Kai Wang\textsuperscript{1,2},
Bingcheng Mao\textsuperscript{1},
Shuai Jia\textsuperscript{1,2},
Yujie Ding\textsuperscript{1,2},
Dongming Han\textsuperscript{1,2},
Tianyi Ma\textsuperscript{1},
Bin Cao\textsuperscript{3}}
\IEEEauthorblockA{\textsuperscript{1}HiThink Research, Hangzhou, China}
\IEEEauthorblockA{\textsuperscript {2}Zhejiang University, Hangzhou, China}
\IEEEauthorblockA{\textsuperscript{3}Zhejiang University of Technology, Hangzhou, China}
\IEEEauthorblockA{\{wangkai7,
maobingcheng, jiashuai, dingyujie2,
handongming2, matianyi\}@myhexin.com, bincao@zjut.edu.cn}
}


\maketitle

\begin{abstract}
Automating code review with Large Language Models (LLMs) shows immense promise, yet practical adoption is hampered by their lack of reliability, context-awareness, and control. To address this, we propose Specification-Grounded Code Review (SGCR), a framework that grounds LLMs in human-authored specifications to produce trustworthy and relevant feedback. SGCR features a novel dual-pathway architecture: an explicit path ensures deterministic compliance with predefined rules derived from these specifications, while an implicit path heuristically discovers and verifies issues beyond those rules. Deployed in a live industrial environment at HiThink Research, SGCR's suggestions achieved a 42\% developer adoption rate—a 90.9\% relative improvement over a baseline LLM (22\%). Our work demonstrates that specification-grounding is a powerful paradigm for bridging the gap between the generative power of LLMs and the rigorous reliability demands of software engineering.
\end{abstract}


\section{Introduction}

Code review, a critical practice in modern software engineering, is instrumental for enhancing software quality, disseminating knowledge, and ensuring architectural consistency \cite{bacchelli2013expectations, sadowski2018modern}. Empirical evidence confirms that reviewed code exhibits significantly lower defect rates \cite{mcintosh2016empirical}. However, the reliance on manual peer examination imposes a substantial burden on developers, creating significant efficiency bottlenecks, especially for complex changes that demand deep contextual understanding \cite{guo2024exploring}. While traditional static analysis tools can automate the detection of superficial issues, they fundamentally lack the capability to comprehend business logic, architectural designs, or complex performance considerations, leaving a vast quality assurance gap that remains manually intensive \cite{li2024enhancing}.

The advent of Large Language Models (LLMs) presents a transformative opportunity for automating code review \cite{yang2024survey,pornprasit2024fine}. Pre-trained on vast code corpora, LLMs demonstrate a remarkable ability to understand programming semantics, identify subtle bugs, and even suggest improvements for logic and design, far surpassing the scope of traditional static analyzers \cite{zhang2023unifying, yu2024fine}. By automating the identification of a wide range of potential issues, LLMs promise to offload significant work from human reviewers, allowing them to focus on high-level strategic decisions and thereby accelerating the development lifecycle \cite{jiang2024survey,fang2024large}.

Despite this promise, the practical application of general-purpose LLMs for code review in production environments is fraught with significant challenges that undermine their reliability and trustworthiness. These challenges collectively hinder their widespread adoption:
\begin{itemize}
    \item \textbf{Lack of Domain-Specific Context:} LLMs trained on open-source data often fail to grasp enterprise-specific business logic, proprietary APIs, and established architectural conventions, leading to irrelevant or incorrect suggestions \cite{xie2024efficient}.
    \item \textbf{Stochasticity and Inconsistency:} The probabilistic nature of LLMs means they can produce different, sometimes conflicting, feedback for the same piece of code, eroding the reproducibility and credibility of the review process \cite{chang2024survey}.
    \item \textbf{Poor Controllability:} LLMs may fixate on trivial stylistic matters while overlooking critical defects in security or performance. They lack the awareness of project priorities to triage issues effectively \cite{bi2024iterative}.
    \item \textbf{Lack of Explainability:} The black-box nature of LLMs makes it difficult for developers to understand the rationale behind a suggestion, fostering distrust and impeding adoption, particularly in mission-critical systems \cite{yu2023temporal}.
\end{itemize}

To bridge this gap between the potential of LLMs and the practical requirements of enterprise-grade software development, this paper introduces Specification-Grounded Code Review (SGCR), a novel framework designed to enhance the reliability, controllability, and trustworthiness of LLM-based code review. The core innovation of SGCR is to ground the LLM's reasoning process in a set of explicit, human-authored specifications. These specifications, encompassing everything from coding standards and security requirements to domain-specific business rules, serve as an authoritative source of truth. By compelling the LLM to operate within these constraints, SGCR directly addresses the aforementioned challenges: it injects necessary domain context, ensures review outputs are consistent and controllable, and improves explainability by linking every suggestion back to a concrete rule. Furthermore, our approach features a novel dual-pathway architecture that synergistically combines deterministic, specification-guided analysis with heuristic, LLM-native issue discovery, ensuring both rigorous compliance and broad coverage.

The main contributions of this paper are threefold:
\begin{enumerate}
    \item \textbf{A Novel Framework:} We propose Specification-Grounded Code Review (SGCR), a new paradigm that grounds LLM-based code review in explicit, project-specific specifications to enhance reliability and trustworthiness.
    \item \textbf{A Dual-Pathway Methodology:} We design and implement a comprehensive methodology featuring two complementary components: \textit{Explicit Specification Injection} for deterministic rule enforcement and \textit{Implicit Specification Discovery} for uncovering emergent issues, balancing rigorous control with heuristic exploration.
    \item \textbf{Empirical Validation in Production:} We deploy and evaluate SGCR in a real-world online business environment at HiThink Research. Our results demonstrate that SGCR achieves a 42\% adoption rate for its suggestions, a relative increase of 90.9\% over a baseline LLM approach, confirming its practical effectiveness and developer acceptance.
\end{enumerate}

\section{Related Work}

Code review, a critical practice for software quality assurance, originated in the 1970s when Fagan formalized structured code inspections based on line-by-line group reviews \cite{fagan2011design}. Modern code review has evolved into a collaborative practice where developers evaluate proposed changes before integration into the main codebase \cite{al2020workload,rahman2016correct}.

To reduce manual effort in code review, various automated approaches have emerged \cite{shi2019automatic,li2019deepreview,tufano2021towards}, with automated static analysis tools being the most widely adopted. These tools identify potential issues—bugs, syntax violations, and best practice deviations—without code execution \cite{ernst2015measure}. Popular tools like SonarQube \cite{campbell2013sonarqube} and PMD \cite{gordon2000pmd} use rule-based mechanisms to flag issues early, supporting quality assurance and coding standards. However, static analysis has limited capability, automatically detecting only up to 16\% of issues found by human reviewers \cite{singh2017evaluating}, indicating both potential for workload reduction and significant coverage gaps.



The recent emergence of LLMs has opened new frontiers for Automated Code Review (ACR) \cite{cihan2024automated,rasheed2024ai}, with researchers leveraging their semantic understanding and generative capabilities to assist the review process \cite{sun2025bitsai}. Lu et al. \cite{lu2023llama} presented LLaMA-Reviewer, which uses LoRA fine-tuning to support all three ACR stages without full model retraining, while Guo et al. \cite{guo2024exploring} found that ChatGPT outperformed the specialized tool CodeReviewer in refinement tasks but underperformed in comment generation. Additionally, Yu et al. \cite{yu2024fine} explored how fine-tuned LLMs can generate more readable review comments and provide relevant explanations for detected issues. The application of LLMs in this domain also overlaps with automated program repair, as Zubair et al. \cite{zubair2025use} examined the use of LLMs to identify and suggest fixes for bugs.

While these advances demonstrate the promising potential of LLMs in automated code review, significant challenges remain in achieving human-level review quality and comprehensive issue detection across diverse codebases and programming contexts.

\begin{figure*}[!t]
	\centerline{\includegraphics[width=\textwidth]{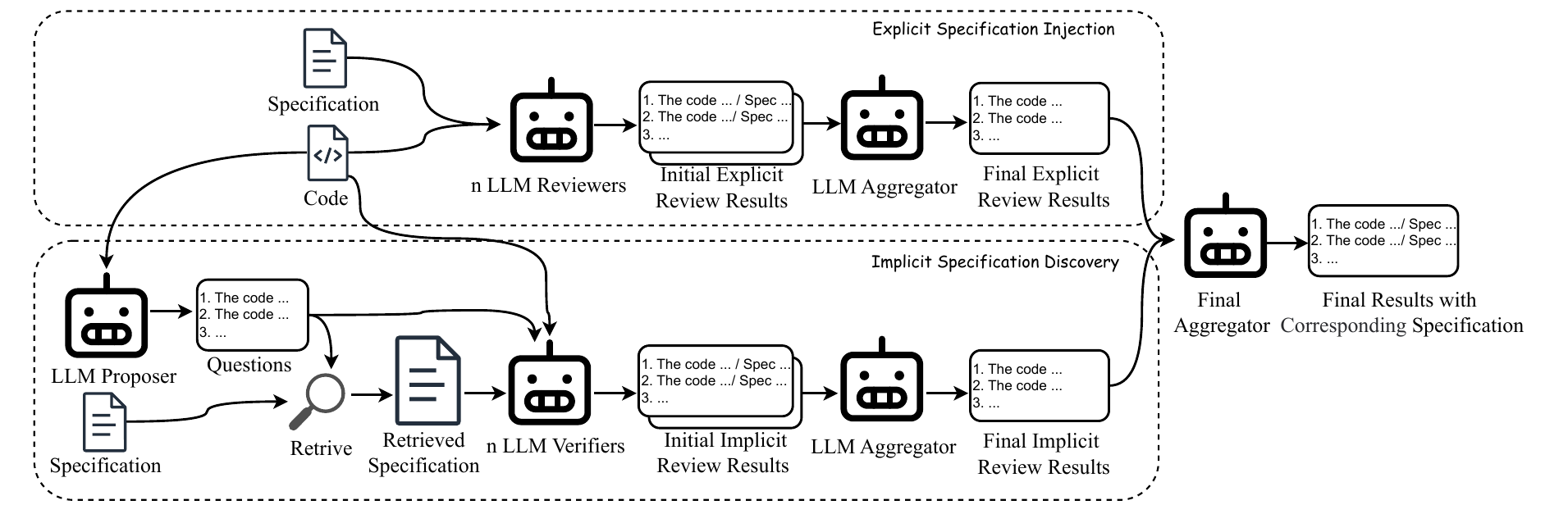}}
	\caption{Overview of proposed SGCR framework. }
	\label{fig:framework}

\end{figure*}

\section{Approach}
As shown in Figure \ref{fig:framework}, the SGCR framework employs a dual-track detection strategy with two core modules: Explicit Specification Injection and Implicit Specification Discovery. The former uses manually annotated specifications as detection benchmarks, while the latter leverages Retrieval-Augmented Generation to automatically discover implicit code constraints. Together, they enable comprehensive identification of potential code issues.


\subsection{Explicit Specification Injection}

The Explicit Specification Injection module shown in Figure \ref{fig:framework} forms the deterministic foundation of the SGCR framework. Its primary objective is to ground the code review process in a set of pre-defined, human-verified rules and standards. This directly addresses the challenges of insufficient domain-specific context and lack of controllability inherent in general-purpose LLMs. By explicitly providing the model with authoritative specifications, we constrain its analytical focus, compelling it to evaluate code against precise, project-relevant quality benchmarks rather than relying solely on its generalized, pre-trained knowledge.

To mitigate the inherent stochasticity of LLM outputs and enhance the reliability of the review, this module employs a parallelized ensemble inference strategy. Instead of a single query, the input source code $C$ and the full set of explicit specifications $S$ are simultaneously fed into multiple, independent LLM instances $\{M_1, M_2, \dots, M_n\}$. Each model instance $M_i$ performs a complete review, generating a candidate initial review result $R_i$. This process can be formalized as:
$$R_i = M_i(C, S), \quad \text{for } i=1, 2, \dots, n$$
This ensemble approach allows for the exploration of diverse reasoning paths, effectively creating a portfolio of potential review comments $\{R_1, R_2, \dots, R_n\}$. This diversity is critical for filtering out anomalous or low-confidence suggestions and identifying a robust consensus among the different model instances.

Subsequently, to synthesize a single, authoritative review from the ensemble's outputs, we introduce a second-order aggregator LLM, denoted as $M_{agg}$. This aggregator model is tasked with a meta-level analysis. It takes the set of candidate results $\{R_1, R_2, \dots, R_n\}$ as its input and performs a comprehensive evaluation, assessing the consistency, relevance, and severity of the issues identified by each base model. The aggregator's function is to merge overlapping suggestions, resolve conflicts, and consolidate the findings into a final, coherent, and actionable review result, $R_{explicit}$. The aggregation process is defined as:
$$R_{explicit} = M_{agg}(\{R_1, R_2, \dots, R_n\})$$
This two-stage process significantly improves the stability and reproducibility of the review results compared to a single-model approach.

In practical business scenarios, it is common to encounter extensive specification documents. Handling such lengthy inputs introduces a critical challenge: the problem of attention dilution. As the input context provided to an LLM grows longer, its ability to maintain a consistent focus and assign appropriate weight to all parts of the text can degrade. This phenomenon may cause the model to overlook or "forget" specifications, particularly those appearing early in a lengthy prompt, thereby compromising the review's thoroughness. To mitigate this, the framework implements a dynamic specification segmentation strategy based on a divide-and-conquer approach. When a specification set $S$ is identified as being excessively long, it is partitioned into smaller chunks $\{s_1, s_2, \dots, s_k\}$. The ensemble review process is then executed in parallel for each chunk against the source code, yielding a set of partial review reports $\{R'_{1}, R'_{2}, \dots, R'_{k}\}$, where each $R'_{j}$ is the aggregated result corresponding to specification chunk $s_j$. A final synthesis step is then performed to integrate these partial reports into a complete and unified final review. This ensures that each specification subset is analyzed within an optimal context, guaranteeing every rule receives sufficient model attention and enabling the system to scale effectively without loss of detail.

\subsection{Implicit Specification Discovery}

While the Explicit Specification Injection module is highly effective for direct compliance checking, its methodology can inadvertently suppress the model's autonomous analytical capabilities. Providing explicit specifications upfront may cause the LLM to overly fixate on the given constraints, adopting a 'checklist' approach that narrows its analytical focus. The critical issue is not that relevant rules are absent from the specification library; rather, the model may fail to identify valid problems—even those with corresponding documented specifications—because its natural reasoning path has been constrained.

To counteract this cognitive fixation and better leverage the model's intrinsic analytical power, the Implicit Specification Discovery module is introduced as a complementary pathway. Its purpose is to decouple the initial problem-finding phase from explicit guidance, allowing the model to first perform an unconstrained, bottom-up analysis of the code. By tasking the model to autonomously identify potential issues first, and only then retrieving relevant specifications for verification, we can surface a class of problems that might have been overlooked by the direct-injection method. This module follows a two-stage propose-and-verify methodology, ensuring that the discovered issues are both a product of the model's deep reasoning capabilities and are ultimately grounded in the project's established standards.

As shown in Figure \ref{fig:framework}, the first stage is unconstrained heuristic analysis. A proposer LLM, denoted as $M_{proposer}$, is tasked with performing a free-form review of the source code $C$ without access to the explicit specification set $S$. The model is prompted to identify potential areas for improvement, logical inconsistencies, or deviations from common best practices based entirely on its internal knowledge base. This process generates a set of hypothetical questions or potential issues, $Q_{hypo}$:
$$Q_{hypo} = M_{proposer}(C)$$
This initial step maximizes the creative and analytical potential of the LLM, allowing it to flag subtle problems that human reviewers or explicit rules might overlook.

The second stage is specification-grounded verification. The raw, hypothetical issues in $Q_{hypo}$ lack the necessary grounding in project context and may contain hallucinations. To refine these hypotheses, we first employ a retrieval mechanism. For each hypothetical issue $q \in Q_{hypo}$, we generate a high-dimensional vector embedding, $\vec{v}_q$. This vector is then used to perform a semantic similarity search against a pre-indexed vector database of all explicit specifications. This retrieves a set of relevant specifications, $S_{retrieved}(q)$, that are semantically related to the potential issue. This step serves to ground the abstract hypothesis with concrete, relevant rules.

Finally, $\{M_{verify,1}, \dots,  M_{verify,n}\}$, a verifier LLM ensemble, is invoked to make a definitive judgment. Each verifier model $M_{verify,i}$ receives the source code $C$, the original hypothetical issue $q$, and the retrieved specifications $S_{retrieved}(q)$ as input. Its task is to confirm whether $q$ constitutes a valid issue in the given context. Mirroring the robust design of the explicit injection module, we aggregate the outputs from the verifier ensemble to produce a final, high-confidence set of implicitly discovered review results, $R_{implicit}$. This structured verification process filters out speculative or irrelevant findings from the initial heuristic analysis, ensuring that the final output of this module is both insightful and trustworthy.

\subsection{Result Aggregation and Optional Code Generation}

The final stage of the SGCR framework is to synthesize the findings from both the explicit and implicit pathways into a single, unified review report. This module integrates the outputs to produce a holistic, de-duplicated, and prioritized final assessment. Based on this final report, it can then optionally generate specification-compliant code suggestions.

First, the results from the Explicit Specification Injection module ($R_{explicit}$) and the Implicit Specification Discovery module ($R_{implicit}$) are fed into a finale aggregation process. This process involves a sophisticated analysis to:
1).  De-duplicate: Identify and merge issues that may have been detected by both pathways.
2).  Prioritize: Rank the identified issues based on severity, typically inferred from metadata within the specifications (e.g., security rules ranked higher than stylistic ones).
3).  Consolidate: Synthesize all verified findings into a single, coherent, and human-readable final review report, $R_{final}$.

This unified report, $R_{final}$, represents the framework's primary output—a reliable and comprehensive assessment of code quality that synergistically combines the strengths of deterministic rule conformance and intelligent, heuristic-based discovery.

Building upon this final review, SGCR also offers an optional capability for specification-guided code generation. For each confirmed issue within $R_{final}$, the system can be prompted to generate a suggested code modification. This generation process is not unconstrained; it is directly guided by the very specifications used to identify the issue:
\begin{itemize}
\item For an issue from the explicit module, its corresponding specification $s \in S$ is used as a direct constraint for the code generation LLM.
\item  For an issue from the implicit module, the set of retrieved specifications $S_{retrieved}(q)$ provides the necessary context and constraints.
\end{itemize}
By leveraging this optional feature, developers can receive concrete and compliant code patches. This significantly reduces the cognitive load and time required for remediation, thereby accelerating the development lifecycle by closing the loop from issue detection to potential resolution.

\section{evaluation}

\subsection{Experimental Setup}
We conduct extensive experiments to evaluate SGCR in production online environments. Specifically, our experiments seek to answer the following research questions:
\begin{itemize}
\item \textbf{RQ1:} How effective is SGCR in improving code review adoption rates in real-world development environments?
\item  \textbf{RQ2:} What is the individual contribution of each component in the SGCR framework to the overall adoption rate?
\item  \textbf{RQ3:} How do developers perceive and accept the SGCR approach in terms of review quality and trustworthiness?
\end{itemize}

\begin{table}[t!]
\centering
\caption{Comparison of Adoption Rates between SGCR and the Baseline.}
 \resizebox{\linewidth}{!}{
\label{tab:rq1_adoption_rate}
\begin{tabular}{lcccc}
\toprule
\textbf{Method} & \textbf{Cases} & \textbf{Adopted} & \textbf{Not Adopted} & \textbf{Adoption Rate} \\
\midrule
SGCR & 279 & 117 & 162 & \textbf{42\%} \\
Baseline  & 248 & 55 & 193 & 22\% \\
\bottomrule
\end{tabular}
}

\end{table}

We evaluate SGCR in a real-world production environment integrated with Java development workflows at HiThink Research. The specification library consists of 140 Java-specific rules authored by experienced developers based on actual business scenarios and development practices, covering code quality standards, security requirements, performance guidelines, and domain-specific business logic. To ensure data privacy and prevent information leakage, we deploy our own model instances: a 32B-parameter LLM serves as the core language model for all review components, while a specialized embedding model provides embedding capabilities for semantic retrieval in the specification discovery module. We measure adoption rate as the primary metric, tracking the percentage of SGCR-generated suggestions that developers choose to implement, and compare against a baseline of vanilla LLM approaches without specification guidance.

\subsection{RQ1: Effectiveness of SGCR}
To address our first research question (\textbf{RQ1}), we conducted a comparative study with 200 participants to evaluate the effectiveness of our proposed SGCR framework against a standard LLM-based code review approach. The primary metric for this evaluation is the adoption rate, which we define as the percentage of review comments generated by the system that are subsequently accepted and implemented by the developer in the final code commit. This metric serves as a strong proxy for the quality, relevance, and trustworthiness of the review suggestions.

Our experiment was conducted on a live production Java project. The baseline approach utilized the same core LLM, but without the specification-grounded mechanisms of SGCR. It performed code reviews based solely on its pre-trained knowledge, akin to many general-purpose LLM code review tools. The SGCR framework leveraged our curated library of 140 specifications. All generated comments from both systems were presented to developers through the standard pull request interface. The results of this comparative analysis are summarized in Table \ref{tab:rq1_adoption_rate}.

As the data clearly indicates, the SGCR framework achieved an adoption rate of 42\%, a substantial improvement over the baseline's 22\%. This represents a relative increase of 90.9\% in developer acceptance of the automated review comments. This significant difference underscores the primary contribution of our work. The baseline model, while capable, often produced generic suggestions (e.g., minor stylistic changes) or occasionally "hallucinated" issues that were not relevant in the context of our business logic.

In contrast, SGCR's suggestions are grounded in the explicit specification library. This grounding mechanism ensures that the review comments are highly contextualized, directly relevant to the project's quality standards, and aligned with domain-specific requirements. By forcing the LLM to reason within the constraints of established rules, SGCR effectively filters out irrelevant noise and enhances the precision and trustworthiness of its output. The higher adoption rate is a direct consequence of developers perceiving the suggestions as more accurate, insightful, and aligned with project goals.

\subsection{RQ2: Ablation Study of SGCR Components}
To answer our second research question (\textbf{RQ2}), we performed an ablation study to dissect the SGCR framework and quantify the individual contributions of its core components: the Explicit Specification Injection module and the Implicit Specification Discovery module. By selectively disabling these components, we can isolate their impact on the overall performance. We evaluated the following configurations:
\begin{itemize}
    \item \textbf{SGCR (Full Framework)}: The complete system with both modules enabled.
    \item \textbf{Explicit-Only}: The framework with only the explicit injection module active. The LLM is guided by specifications but does not perform unconstrained analysis.
    \item \textbf{Implicit-Only}: The framework with only the implicit discovery module. The LLM first proposes issues heuristically, which are then verified against the specification database.
    \item \textbf{Baseline}: The vanilla model with no SGCR components.
\end{itemize}

\begin{table}[t!]
\centering
\caption{Ablation Study of SGCR Components on Adoption Rate.}
 \resizebox{\linewidth}{!}{
\label{tab:rq2_ablation}
\begin{tabular}{lc}
\toprule
\textbf{Configuration} & \textbf{Adoption Rate (\%)} \\
\midrule
\textbf{SGCR (Full Framework)} & \textbf{42\%} \\
\quad \textit{w/o} Implicit Specification Discovery (Explicit-Only) & 37\% \\
\quad \textit{w/o} Explicit Specification Injection (Implicit-Only) & 29\% \\
Baseline (No SGCR modules) & 22\% \\
\bottomrule
\end{tabular}
}

\end{table}

The results of the ablation study, measured by adoption rate, are presented in Table \ref{tab:rq2_ablation}. The findings from the ablation study are highly informative. First, the Explicit-Only configuration achieves an adoption rate of 37\%. This represents the most significant performance leap from the 22\% baseline, demonstrating that explicitly guiding the LLM with project-specific rules is the most critical factor for improving review quality. This directly addresses the problem of LLMs lacking domain context.

Second, the Implicit-Only configuration yields a 29\% adoption rate. While more effective than the baseline, its performance is lower than the Explicit-Only version. This is expected, as the initial unconstrained analysis can still generate speculative or lower-priority suggestions. However, the improvement over the baseline confirms the value of its propose-and-verify mechanism, which grounds the LLM's heuristic findings in established specifications, thereby filtering out hallucinations and increasing relevance.

Finally, the full SGCR framework, which synergistically combines both modules, achieves the highest adoption rate of 42\%. The 5-percentage-point increase from the Explicit-Only configuration highlights the complementary nature of the implicit discovery module. It acts as a valuable "safety net," uncovering valid issues that might be missed when the model is strictly constrained by the explicit specification set. This demonstrates that the two pathways work in concert to achieve a review process that is both disciplined and discovery-oriented.

\subsection{RQ3: Developer Perception and Qualitative Feedback}

To assess developer acceptance and perception of SGCR, we conducted a qualitative study with 15 developers using anonymous surveys and semi-structured interviews. The feedback was overwhelmingly positive, highlighting three primary benefits of the framework.

First, developers reported enhanced trust and explainability, as SGCR's suggestions are directly linked to specific, human-authored rules, in contrast to the "black-box" nature of baseline models. Second, they noted increased relevance and reduced noise, confirming that SGCR effectively filters out generic or irrelevant comments and focuses on issues pertinent to the project's specific business logic and architectural standards. Finally, developers, particularly junior ones, recognized the system's educational value, as it provided just-in-time learning opportunities for project-specific best practices.

Constructive feedback was also collected for future work. The main concerns included the maintenance overhead of the specification library and a desire for more interactive features, such as the ability to ask for clarification on a suggestion. These insights provide a clear direction for the future evolution of the SGCR framework.
\section{Conclusion}

To overcome the critical reliability and context-awareness failures of LLMs in code review, this paper introduces the Specification-Grounded Code Review (SGCR), a framework that grounds the model's analysis in project-specific rules using a novel dual-pathway architecture. Our extensive evaluation in a real-world production environment demonstrates that SGCR achieves a 42\% adoption rate for review suggestions, representing a 90.9\% relative improvement over baseline LLM approaches, thereby establishing specification-grounded guidance as an effective strategy for bridging the gap between AI potential and enterprise requirements in automated code review systems.

\bibliographystyle{IEEEtran}
\bibliography{IEEEexample}

\end{document}